\documentclass[10pt,journal,compsoc]{IEEEtran}

\ifCLASSOPTIONcompsoc
  \usepackage[nocompress]{cite}
\else
  \usepackage{cite}
\fi

\usepackage{setspace,hyperref}

\usepackage{amssymb,amsmath,color}

\usepackage{graphicx}
\usepackage{epstopdf}

\definecolor{dgreen}{rgb}{0,0.6,0}

\definecolor{bred}{rgb}{0.7,0,0}

\begin{document}

\title{On the Performance and Implementation of Parallax free Video See-Through Displays}

\author{Ricardo~Augusto~Borsoi,
and~Guilherme~Holsbach~Costa%
\IEEEcompsocitemizethanks{\IEEEcompsocthanksitem R.A. Borsoi is with the Department of Electrical Engineering, Federal University of Santa Catarina, Florian\'opolis, Brazil.%
~E-mail: raborsoi@ucs.br
\IEEEcompsocthanksitem G.H. Costa is with the Department of Mechanical Engineering, University of Caxias do Sul, Caxias do Sul, Brazil.%
~E-mail: holsbach@ieee.org}%
\thanks{Manuscript received xx xx, xxxx; revised xx xx, xxxx.}}


\IEEEtitleabstractindextext{%
\begin{abstract}
In see-through systems an observer watches a (background) scene partially occluded by a display. In this display, usually positioned close to the observer, a region of the background scene is shown, yielding the sensation that the display is transparent. To achieve the transparency effect, it is very important to compensate the parallax error and other distortions caused by the image acquisition system. In this paper a detailed study of a video see-through methodology with parallax correction is performed. In a system composed by two cameras --- one directed to the user and another to the background scene --- and a display, the relative position between the user, the display and the scene is estimated using a feature detection algorithm and the parallax error is compensated assuming a planar scene model. The application of the proposed methodology on Driver Assistance Systems (DAS) is proposed. A theoretical assessment of the algorithm shows that although approximations are proposed to simplify the methodology and reduce the computational cost, such as the planar scene model and fixed working distance, on some practical situations their effects can be neglected without noticeable impact on the perceptual quality of the solution.
\end{abstract}

}

\maketitle

\IEEEdisplaynontitleabstractindextext

\IEEEpeerreviewmaketitle

\ifCLASSOPTIONcompsoc
\IEEEraisesectionheading{\section{Introduction}\label{INTRO}}
\else
\section{Introduction}
\label{INTRO}
\fi

\IEEEPARstart{A}{ugmented} reality (AR) is a prominent technology which can enhance or change the perception of real environments by merging real and virtual objects. This technology allows new degrees of interaction by displaying information that could not be directly perceived by the user's senses \cite{Azuma97,Krevelen10}. Recently, AR technologies have begun to move out of laboratories and research centres around the world right into industries and consumer markets of several kinds, providing a new type of human-machine interaction with a large degree of intuitiveness and user friendliness~\cite{Krevelen10}.

AR has several applications, including consumer electronics products, medical visualisation using non-invasive imaging sensors during surgery \cite{Azuma97}, assistance on the assemblage or maintenance of complex machinery, military aircraft exhibition of flight information or target aiming display \cite{Azuma97} and the presentation of traffic information for drivers on intelligent vehicles \cite{Gavrila01}. Among these applications, vehicular Driver Assistance Systems (DAS) are recently attracting a great deal of interest from the community. DAS are designed for safety and better driving and can reduce the large number of injuries associated with traffic accidents, providing the driver information that allows safer manoeuvres, for example. Moreover, complex DAS are becoming increasingly available to the average customer, motivating the development of new AR solutions \cite{Gavrila01,Kucukay04}. Considering this, one of the most promising applications of AR is the so called see-through system, which consists in revealing content hidden in a blind spot by showing information on a display while giving the user the impression that it is a transparent screen. With this kind of AR based DAS, a driver can, for example, see what is behind the pillars of the vehicle or receive relevant information on the windshield without occluding his view, providing increased safety and comfort \cite{Kucukay04}, see through the fog or in poor sighting conditions  \cite{Gavrila01,Rosler06} using a monitor displaying images acquired by special (infrared) cameras, or see through obstructing vehicles in a lane while performing a passing manoeuvre~\cite{Gomes2012passing}.

The most simple and straightforward way to achieve the see-through effect is by using transparent displays and superimposing the additional information. This approach is called optical see-through \cite{Rolland00}. Although presenting a simple design, optical see-through systems require specific displays and may not be suitable for situations such as a transparent display over solid obstacles. Examples of such systems can be seen in \cite{Lee13}, where a semi-transparent monitor was used to provide the user interaction with the 3D space behind the screen through hand gestures, and in \cite{Hilliges12}, where a system allows the user to interact with virtual objects through a transparent table.

A more generic approach considers opaque displays. In this case, background scene images have to be acquired, processed and projected into the display in order to achieve a virtual transparency effect. Notice that in order to effectively achieve virtual transparency and consequently enhance the perception of the environment in AR applications, see-through systems require a seamless merging of the virtual information shown in the display with the real background scene seen by the user.
This provides the integration of the displayed image with its surrounding context, which is important for many AR related tasks~\cite{baudisch2002keepingContext,Pucihar14}.
However, matching virtual and real world is probably the major problem in virtual transparency. The digital images, captured by the cameras of a see-through system, usually have a spatial perspective different from that of the user, introducing a parallax error between the would-be-seen image (in the user's occluded field of view, for example) and the effectively acquired image. Parallax compensation is not an easy task, and avoiding to delegate this to the system ends up limiting the range of see-through applications, since perceptible perspective distortions are left in the displayed image~\cite{steinicke2011realisticPerpective}. Mainly because of parallax problems, most video see-through systems proposed up to date have been restricted to Head Mounted Displays (HMD), where the field of view of the camera is rigidly aligned with the user's \cite{Hill11,Rolland00}.

Considering the use of opaque displays, many methods have been proposed recently for the correction of the parallax error, most of which focus on a planar geometry approximation of the system comprised by the scene, obstacle and user. This allows the employment of a simple transformation consisting of a combination of rotation, translation and scaling over the image acquired by the camera in order to align it with the user's perspective \cite{Hill11,Tomioka13,Tomioka14,Pucihar13,Matsuda13,Uchida13}. Although the approximation of the scene geometry by a plane provides a solution with low computational cost, it is widely known that distortions are usually present in the processed image due to non-idealities such as occlusion. This motivated the consideration of a complex 3D scene modelling and reconstruction which was employed in some works, albeit at the expense of a high computational cost \cite{Baricevic12,barivcevic2014user,Unuma14}.

Despite the number of papers focusing on the parallax correction available on the literature, in most of them the proposed methods are not characterized in detail. For instance, the works \cite{Hill11,Baricevic12,barivcevic2014user,Unuma14,Tomioka14} do not offer any mathematical description of their methods. Although different methods and implementations are proposed in the remaining works, the formal or mathematical descriptions provided are mainly brief and an in-depth evaluation of the behavior and implementation of the method is not provided, since most works are instead focused towards the rendering of virtual objects or the subjective quality assessments over implemented prototypes.
Furthermore, the performance of the methods was not evaluated quantitatively in the previous works, which motivates the study of the impact of hypotheses about the scene distance and geometry on the quality of the resulting images, as well as the assessment of the actual benefit of the more costly and elaborated methods recently adopted for the scene distance estimation.

For the case of optical see-through HMDs, a detailed theoretical assessment of its performance has proven to be valuable to determine both the sources of errors and their effect in the alignment of the projected virtual objects with the real scene~\cite{Holloway1995registrationErrorsAR}.
This provides insight into the calibration of the system, and in some cases may allow for informed developments or simplifications of the technique.
For instance, considering that the geometry of distant scenes is well approximated by a planar model allows the adoption of simpler motion models in object motion detection~\cite{Irani1998movingDetection}, or the use of simpler algorithms given a minimum working/scene distance for the multi-camera acquisition of wide field of view images~\cite{Swaminathan2000nonmetric}.
Performing a detailed theoretical study of video see-through devices can be a valuable source of insight into the effects of the diverse errors and approximations involved in such systems.
This motivates further developments of the method and can lead to substantiated guidelines for specifying working conditions and choosing parameter values.

In this paper, a parallax free video see-through method similar to those based on \cite{Hill11} is described in detail. In the selected system, images of the occluded areas of the scene are acquired by a frontal camera and the position of the user head is estimated through the use of an user-oriented (back) camera and a face detection algorithm. The displacement between the frontal camera and user's position is calculated and the acquired scene image is cropped according to a planar geometry approximation to correct the parallax error, before being shown in the display. Afterwards the proposed method is then theoretically evaluated to account for the effects of the considered approximations and for the influence of the parameters of the system on the processed image using planar geometry. This allows the evaluation of the indicated working conditions and parameters, and brings important conclusions relating the performance of the method to characteristics of the scene and camera, which in turn leads to the proposition of an approximation that substantially reduces the computational cost for some applications. Finally, the proposed methodology is illustrated in practice through examples implemented in a simulation environment and through a physical prototype.

This paper is divided as follows. In Section \ref{METHOD} the video see-through methodology considered is described in detail, and the method for estimating the user's position is presented. In Section \ref{PARALLAX} the geometric parallax correction algorithm is developed under some approximations for the background scene. The employed approximations are evaluated theoretically in Section \ref{ANALYSIS}, and important conclusions about the behavior of the system are obtained. 
In Section~\ref{SIMUL} computer simulations illustrate the proposed method.
Finally, some concluding remarks end the paper.

\section{Video See-Through Systems} \label{METHOD}

\begin{figure}[htb]
	\centering
		\includegraphics[width=60mm]{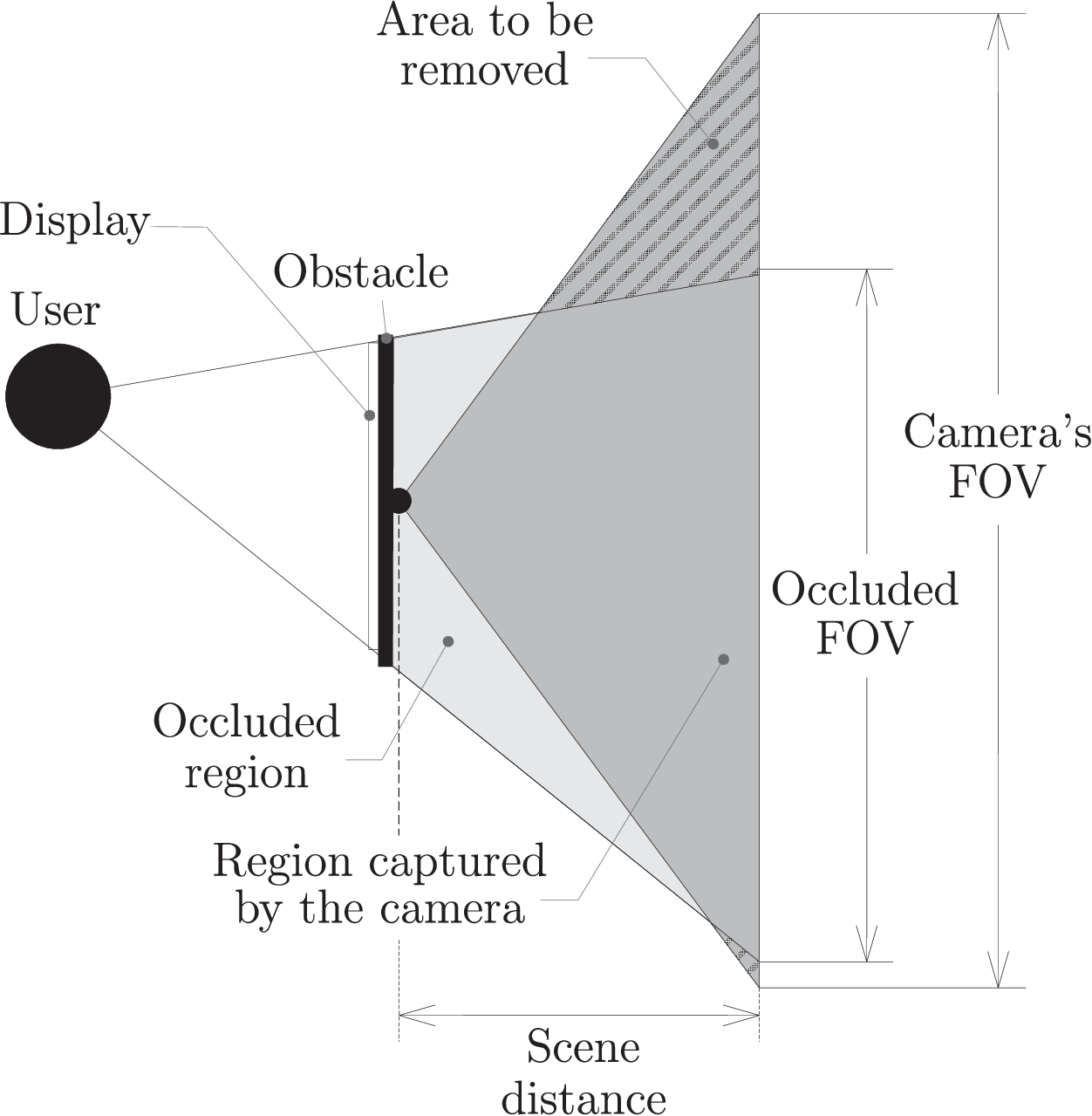}
	\caption{Simplified problem diagram.}
	\label{fig:diagrama}
\end{figure}

A typical video see-through system configuration is shown through a top-view diagram in Figure~\ref{fig:diagrama}, where an user has his field of view partially blocked by an obstacle. A monitor, fixed over the obstacle, shows images acquired by a camera fixed on the other side of this obstacle and turned to the observed scene. Hereafter, this camera will be called \textit{Frontal Camera}. To efficiently merge the image displayed in the monitor with the real world, aiming at a transparency effect, parallax compensation is required in practical applications, since the frontal camera does not have the same field of view (FOV) as the user (see Figure~\ref{fig:diagrama}).

The parallax error depends on the image acquisition system, on the relative position between user and camera, and on the geometry of the captured scene. Since some of this information is unavailable in practice, it must be estimated for parallax correction to be performed.

For the estimation of the user position in the 3-D world, the most common approach is the use of a (single) camera turned to the user for estimating its location based on some sort of feature detection, since it yields low cost hardware implementations, as in smartphones and tablets \cite{Matsuda13,Unuma14}. Hereafter, the camera used for user detection will be called \textit{Back Camera}.

\begin{figure}[!bht]
	\centering
	\includegraphics[width=87mm]{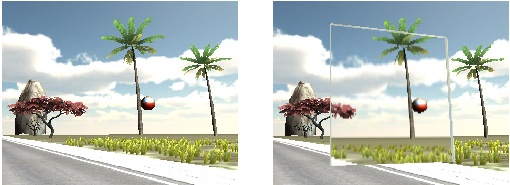}
	\caption{An example of a video see-through system (notice the distortions near the lower border of the obstacle due to the planar scene approximation).}
	\label{fig:illustrative_simul_intro}
\end{figure}

Having estimated the user's position, some method have to be then applied in order to compensate the parallax error. Most algorithms are based on planar geometry, which shall be described in depth on the next section since it is the focus of the present work. An illustrative example of this methodology is depicted through a simulation in Figure~\ref{fig:illustrative_simul_intro}. More complex methods are also employed, displaying a \mbox{3-D} model of the occluded part of the scene created based on the acquired image, intending to address the effects that non-planar geometry have on the parallax error.

\section{Planar Geometry Based Parallax Compensation}\label{PARALLAX}

The employed approach can be divided in two steps. First, an image of the user is acquired by the \textit{Back Camera} and his spatial position is estimated based, for example, on a face detection algorithm. Second, the background (would-be-seen) scene is acquired by the \textit{Frontal Camera}, and the parallax error is then corrected by computing the correspondence between the occluded and camera's field of view using the estimated user's head position. A region of the image acquired by the \textit{Frontal Camera} corresponding to the occluded field of view can then be cropped, resized and finally projected into the display, pursuing a seamless blending with the remaining scene.

The parallax compensation is based on planar geometry, where the obstacle and the background scene are regarded as flat planes on the three dimensional space and the camera and user views are assumed to follow the Pinhole Camera model, according to the diagram in Figure~\ref{fig:diagrama}, which also depicts the main objective of selecting the proper region on the \textit{Frontal Camera}'s image corresponding to the user's occluded field of view, and the dependence this region have with the user's position and working/scene distance. In the following subsections, the aforementioned steps will be treated in further detail.

\subsection{User's head position detection}

In order to estimate and correct the parallax error one needs to know the user's eyes position, in this case using the center of the obstacle as a reference coordinate system. Any movement of the user's viewing position changes the portion of the background scene which should be accordingly viewed through the obstacle, imposing the need for dynamic compensation.

The estimation of the user's view position is generally based on the identification of some facial features whose real dimension are known in advance. Common algorithms that may be employed for this end include, for example, facial detection, eyes detection and pose estimation \cite{Zhang10survey,Murphy09headPose}. With the detected and real feature length at hand, planar geometry can be employed to compute the user position. In describing the position calculation, the width of the head returned by a face detection algorithm is going to be employed as a feature, without loss of generality (different features can be employed following the same procedure).

\begin{figure}[htb]
	\centering
		\includegraphics[width=60mm]{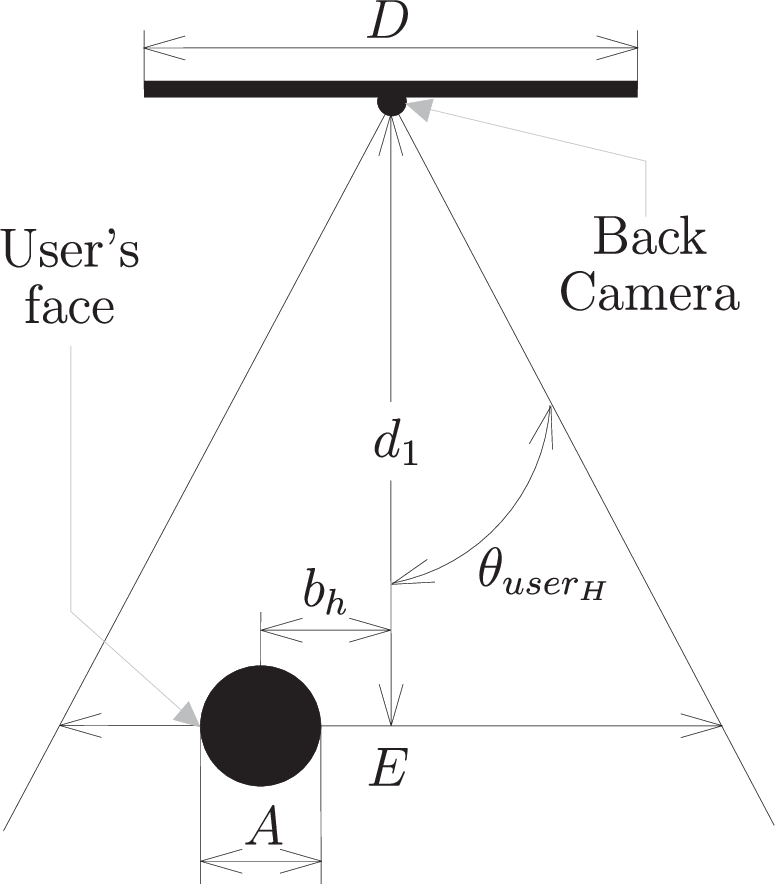}
	\caption{User's head position detection.}
	\label{fig:user_pos_detection}
\end{figure}

A diagram representing the top view of the user-obstacle part of the system can be seen in Figure~\ref{fig:user_pos_detection}. The goal of the following procedure is to compute the distance $d_1$ between the user and the display, and the horizontal distance $b_h$ between the user and the \textit{Back Camera}'s focal axis. In order to maintain the remaining of this work simple, the method will be presented in only two dimensions, considering a top view of the \mbox{3-D} system. The extension for the three dimensional case is trivial and therefore will not be addressed.

To compute $d_1$ and $b_h$, we first define the ratio between the face width $A$ and the scene width $E$ as:
\begin{equation} \label{eq:face1}
    \alpha = \frac{A}{E} = \frac{A_{px}}{E_{px}}
    \,\text{,}
\end{equation}
where $E_{px}$ and $A_{px}$ are the widths of the image and detected face in pixels. This way, using equation \eqref{eq:face1} and trigonometric relations, distance $d_1$ can be written as:
\begin{equation} \label{eq:d1}
    d_1 = \frac{A}{2\alpha \tan(\theta_{user_H})} = \frac{A}{2 \frac{A_{px}}{E_{px}} \tan(\theta_{user_H})}
    \,\text{,}
\end{equation}
where $\theta_{user_H}$ is the angle of view of the \textit{Back Camera} in the horizontal plane. The distance between the user and the focal axis can be similarly computed by first defining the ratio between the distance $b_h$ and the scene width $E$:
\begin{equation} \label{eq:face2}
    \epsilon_h = \frac{b_h}{E} = \frac{b_{h_{px}}}{E_{px}}
    \,\text{,}
\end{equation}
where $b_{h_{px}}$ is the horizontal displacement of the face in pixels. Then, using equation \eqref{eq:face2} together with trigonometric relations gives us the position as:
\begin{equation} \label{eq:bh}
    b_h = 2\epsilon_h d_1 \tan(\theta_{user_H}) = 2\frac{b_{h_{px}}}{E_{px}} d_1 \tan(\theta_{user_H})
    \,\text{.}
\end{equation}
Note that the distances computed by \eqref{eq:d1} and \eqref{eq:bh} are given with respect to the \textit{Back Camera}. Since this camera is usually fixed on the top of the display, its position must be compensated in order to set the coordinate system on the center of the display.

Observing the obtained equations, it can be seen that the estimated user's position depends directly on the real and estimated features $A$, $A_{px}$ and $b_{px}$. Since errors on any of these variables will directly affect the parallax compensation, leading the \textit{Frontal Camera}'s image to be transformed into a viewpoint different than the user's, precise knowledge of $A$, $A_{px}$ and $b_{px}$ is necessary in order to attain a good result.

In principle, the freedom in choosing the type of feature $A$ allows for the choice which proves most advantageous for the application. It is desired, for example, to have the detection $A_{px}$ as precise as possible, or to have minimum variability of $A$ between different users in order to increase the system robustness. However, since real time operation is one of the main requirements of a see-through system, it is necessary to deploy simple and fast algorithms for the feature detection. Unfortunately, this comes at the expense of a reduced quality, which in turn makes it unlikely for the user's position to be precisely known. This incites the need to formally investigate the influence these errors have on the quality of the corrected view, which is one of the objectives of Section~\ref{ANALYSIS}.

\subsection{Parallax Compensation}

\begin{figure}[htb]
	\centering
		\includegraphics[width=60mm]{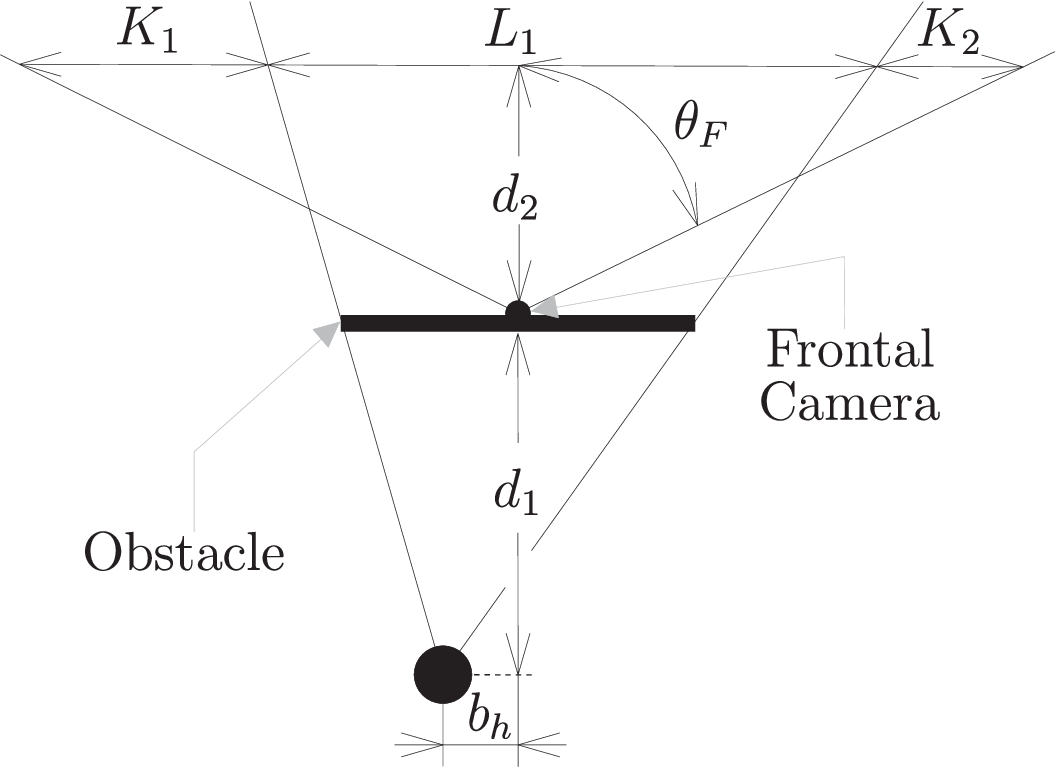}
	\caption{Diagram of the problem with the user close to the \textit{Back Camera}'s focal axis.}
	\label{fig:recorte1}
\end{figure}

\begin{figure}[htb]
	\centering
		\includegraphics[width=60mm]{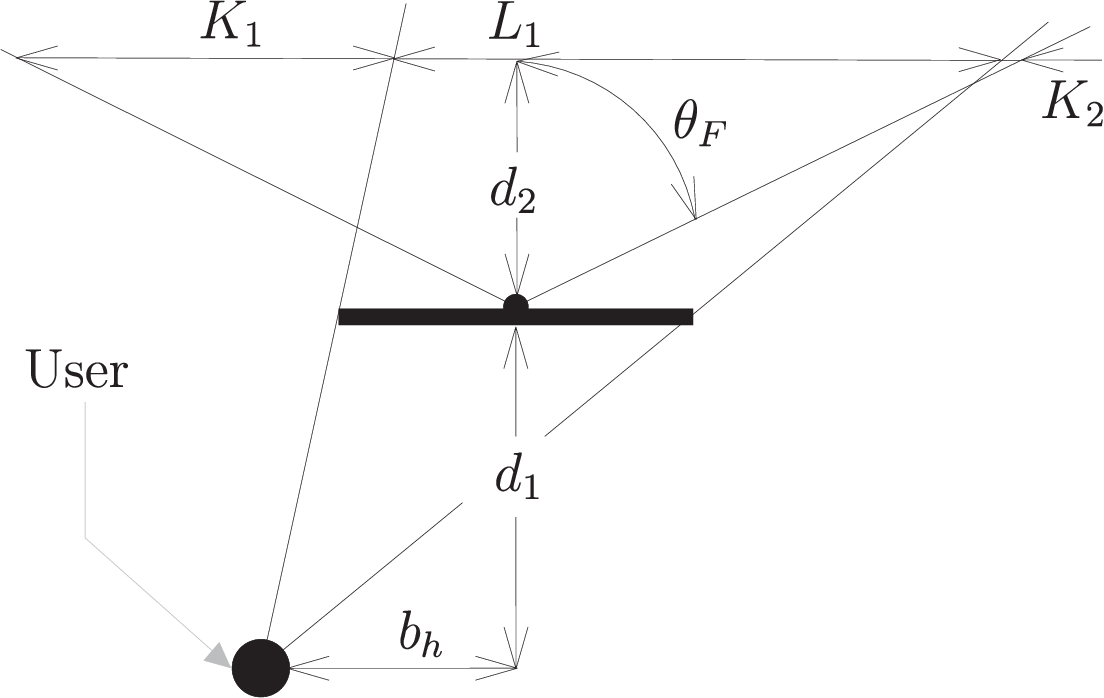}
	\caption{Diagram of the problem with the user far from the \textit{Back Camera}'s focal axis.}
	\label{fig:recorte2}
\end{figure}

The parallax compensation task basically consists in selecting a region of interest in the \textit{Frontal Camera}'s image to show on the display depending on the user's position, aiming to provide the perception of a transparent display. As the present approach is based on planar geometry, the background scene will be approximated by a flat plane, parallel to the obstacle and the display. A diagram of the problem is depicted in Figures~\ref{fig:recorte1} and~\ref{fig:recorte2}. The objective is to compute the areas of the \textit{Frontal Camera}'s image to be discarded and the region of interest, given by $K_1$, $K_2$ and $L_1$.

We first express the regions to be discarded by means of its ratio with respect to the scene width:
\begin{equation} \label{EQ:RATIO}
    R_{\{1,2\}} = \frac{K_{\{1,2\}}}{K_1 + K_2 + L_1}
    \,\text{.}
\end{equation}
From trigonometric relations we have that $K_1 + K_2 + L_1=2d_2\tan(\theta_F)$, with $d_2$ being the distance between the \textit{Frontal Camera} and the scene (usually called {\it working distance)} and $\theta_F$ the \textit{Frontal Camera}'s angle of view. The areas to be discarded are obtained similarly, given the user position and obstacle width $D$:
\begin{equation} \label{eq:paralax2}
\begin{array}{c}
    K_1=d_2\tan(\theta_F)-(d_1+d_2)\displaystyle\frac{0.5D-b_h}{d_1}-b_h   \,\text{,} \\[.3cm]
    K_2=d_2\tan(\theta_F)-(d_1+d_2)\displaystyle\frac{0.5D+b_h}{d_1}+b_h   \,\text{.}
\end{array}
\end{equation}

Applying \eqref{eq:paralax2} in \eqref{EQ:RATIO}, the ratio $R_{\{1,2\}}$ representing the area to be discarded is given by:
\begin{equation} \label{EQ:RATIO1}
    R_{\{1,2\}} = \frac{d_2\tan(\theta_F)-(d_1+d_2)\frac{0.5D\mp b_h}{d_1}\mp b_h}{2d_2\tan(\theta_F)}
    \,\text{.}
\end{equation}

Since the ratio in \eqref{EQ:RATIO1} is a dimensionless quantity, the \textit{Frontal Camera}'s resolution must be considered in order to calculate the number of columns to be discarded on the acquired image, which will be given by:
\begin{equation} \nonumber
    K_{\{1,2\}_{px}} = \left\lfloor R_{\{1,2\}}E_{px}^F \right\rceil
    \,\text{,}
\end{equation}
where $K_{\{1,2\}_{px}}$ is the number of columns to be removed from the borders of the image (in pixels), $E_{px}^F$ is the \textit{Frontal Camera}'s image width in pixels and $\left\lfloor \cdot \right\rceil$ is the nearest integer operator. Like in the previous section, the extension of the method to the third dimension is trivial and therefore will not be addressed.

\section{Theoretical Evaluation} \label{ANALYSIS}

Although the inherent simplicity of planar geometry based parallax correction techniques makes real time application possible, it still imposes limitations that may have significant impact on the performance of the system. The adoption of a planar scene model reduces the algorithm complexity, but the distance $d_2$ from the scene to the obstacle must still be known in advance or estimated, which involves the use of high cost techniques such as LASER scanners \cite{Reiter11}, stereoscopic imaging \cite{Reiter11} or image registration \cite{Chen07}. The working distance estimation constitutes one of the main limitations of planar geometry based systems, as well as another source of errors in the algorithm along with the user's position estimation. Furthermore, the geometry of the scene is hardly a flat plane in real world applications. To make matters worse, previous works are focused on implementation and do not provide any discussion about the impact of these non idealities on the system performance.

In this section we are going to investigate how the aforementioned errors affect the performance of the see-through systems studied by employing differential sensitivity analysis. Sensitivity analysis aims to determine the degree to which a change in an input parameter affects the output of a model, identifying which variables contribute the most to output variability. In fact, it is important to distinguish between two important properties of input variables. A sensitive parameter is that whose small variation yields a significant output variation. An important parameter, on the other hand, besides being sensitive, presents a significant variation itself, such that its effect on the output cannot be neglected \cite{Hamby94}.

Differential sensitivity analysis is based on the sensitivity coefficient $\phi_i$, which is the ratio of change of the model output to the change in one specific input variable, while keeping the remaining ones constant \cite{Hamby94}. This method consists of a linear approximation for the model, needing the input variables to be independent and the parameter disturbances to be small. Nevertheless, its simplicity provides a good understanding of the system's behavior and gives important intuition about its workings.

For the see-through case, the sensitivity ratio of \eqref{EQ:RATIO1} will be evaluated in relation to the user position and to the working distance, with an emphasis to the latter since it will be shown to be an important parameter mainly due to its very high uncertainty originating from inaccurate algorithms, which must necessarily be very fast in order to allow real time operation. Furthermore, it is desired to know how the system is influenced by different working conditions, thus identifying when it is possible to achieve a satisfactory performance and avoid situations under which it won't work accordingly.

Afterwards, the influence of non-planar geometry will be accounted for by modelling it through the phenomenas of occlusion and foreshortening. Albeit relatively simple, these effects encompass most of the geometry based distortions experienced by the user. By studying the distortion behavior it will be possible to evaluate its influence on the see-trough system and its dependence on the remaining variables. Other sources of distortion, such as non-lambertian surfaces, will not be considered in the present study.

\subsection{Sensitivity analysis of $R_{\{1,2\}}$}
In order to measure the sensitivity of the cutting ratio to errors in the estimation of the user's position, we compute its sensitivity coefficient by taking its derivative with respect to $d_1$ and $b_h$:
\begin{equation} \label{EQ:ANALYSIS_USR1}
\phi_{d_1}=\frac{\partial R_{\{1,2\}}}{\partial d_1}=\frac{\mp b_h+0.5D}{2 \tan(\theta_F) d_{1}^2}
\,\text{,}
\end{equation}
\begin{equation} \label{EQ:ANALYSIS_USR2}
\phi_{b_h}=\frac{\partial R_{\{1,2\}}}{\partial b_h}=\pm \frac{1}{2d_1 \tan(\theta_F)}
\,\text{.}
\end{equation}

Examining the relationships between the computed sensitivities and the remaining variables, the first thing that can be noticed is that the sensitivities are independent of the scene distance, leading to a constant error as function of $d_2$. Furthermore, $\phi_{d_1}$ and $\phi_{b_h}$ are inversely proportional to the user-obstacle distance, which means that errors in the user's position estimation have less impact on the system output if he is not very close to the obstacle.
Since both expressions are also divided by $\tan(\theta_F)$, the choice of a larger angle of view $\theta_F$ for the \textit{Frontal Camera} may lead to a significant reduction of errors caused by the estimation of $d_1$ and $b_h$. Besides, it is well known in the literature that larger values of $\theta_F$ also allows for a larger set of admissible user positions.

It is also interesting to point out that the sensitivity $\phi_{d_1}$ varies with $b_h$, as it balances linearly between each side of the display $R_1$ and $R_2$ depending on how distant the user is from its focal axis. The minimum $\phi_{d_1}$ for both sides is achieved when $b_h$ and $D$ are small, corresponding to a small obstacle with the user close to its center, although systems with large displays will be less affected by changes in the position $b_h$ (as long as it respects the maximum admissible value for the occluded view to fall inside the \textit{Frontal Camera}'s field of view).

An important aspect to be evaluated is how the cutting ratio behaves in the presence of errors in the working distance estimation, which in practice is very imprecise. This is given by its sensitivity coefficient which is computed by taking its derivative with respect to $d_2$:
\begin{equation} \label{EQ:ANALYSIS_D2}
    \phi_{d_2}=\frac{\partial R_{\{1,2\}}}{\partial d_2}=\frac{-0.5D}{2d_{2}^2\tan (\theta_F )}
    \,\text{.}
\end{equation}

It can be seen that the sensitivity $\phi_{d_2}$ is inversely proportional to the scene distance, which means that if the scene distance is considerably large, any errors in the estimation of $d_2$ will have little effect on the cutting ratio and consequently on the quality of the system output. Moreover, $\lim_{d_2 \rightarrow \infty}\phi_{d_2}=0$, which means that the cutting ratio becomes independent of the scene distance as it approaches infinity. Furthermore, it can be shown that
\begin{equation} \label{EQ:ANALYSIS_D2_MONOTONE}
    \frac{\partial\phi_{d_2}}{\partial d_2}=\frac{D}{2d_{2}^3\tan (\theta_F )}
    >0\,\text{,} \quad\forall\,\, d_2>0
    \, \text{,}
\end{equation}
which means that $\phi_{d_2}$ is monotonic with respect to $d_2$. This implies that given some fixed value $\bar{d_2}$, the error strictly decreases for larger scene distances (i.e. $\phi_{d_2}(\bar{d_2})>\phi_{d_2}(d_2)$ for $d_2>\bar{d_2}$). This means that if we know some minimum working distance $d_{2\min}$ that the system will be subject to in a given practical implementation, the scene distance estimation may possibly be completely dropped by setting a fixed $d_2\rightarrow\infty$ if the corresponding errors, which are bounded by $\phi_{d_2}(d_{2\min})$, are acceptable for the application. This leads to significant computational savings when the typical values for $d_2$ are large.

It is important to note the influence of the \textit{Frontal Camera}'s angle of view $\theta_F$ in the sensitivity $\phi_{d_2}$. Since the expressions on \eqref{EQ:ANALYSIS_D2} and \eqref{EQ:ANALYSIS_D2_MONOTONE} are being divided by $\tan(\theta_F)$, a larger angle of view significantly reduces the sensitivity of the system to errors in the scene distance estimation. Furthermore, the rate at which the sensitivity approaches zero is also higher, implying that for larger $\theta_F$ the minimum scene distance $d_{2\min}$ at which an acceptable error threshold is reached is smaller, making the approximation of $d_2\rightarrow\infty$ valid for a wider range of practical scenarios. This again reinforces the importance of a wide field of view for the \textit{Frontal Camera}.
It is also interesting to note that smaller obstacles sizes $D$ make the system less sensible to errors in the scene distance estimation.

\subsection{Effect of Non-Planar Geometry}

\begin{figure}[ht]
\centering
\includegraphics[width=80mm]{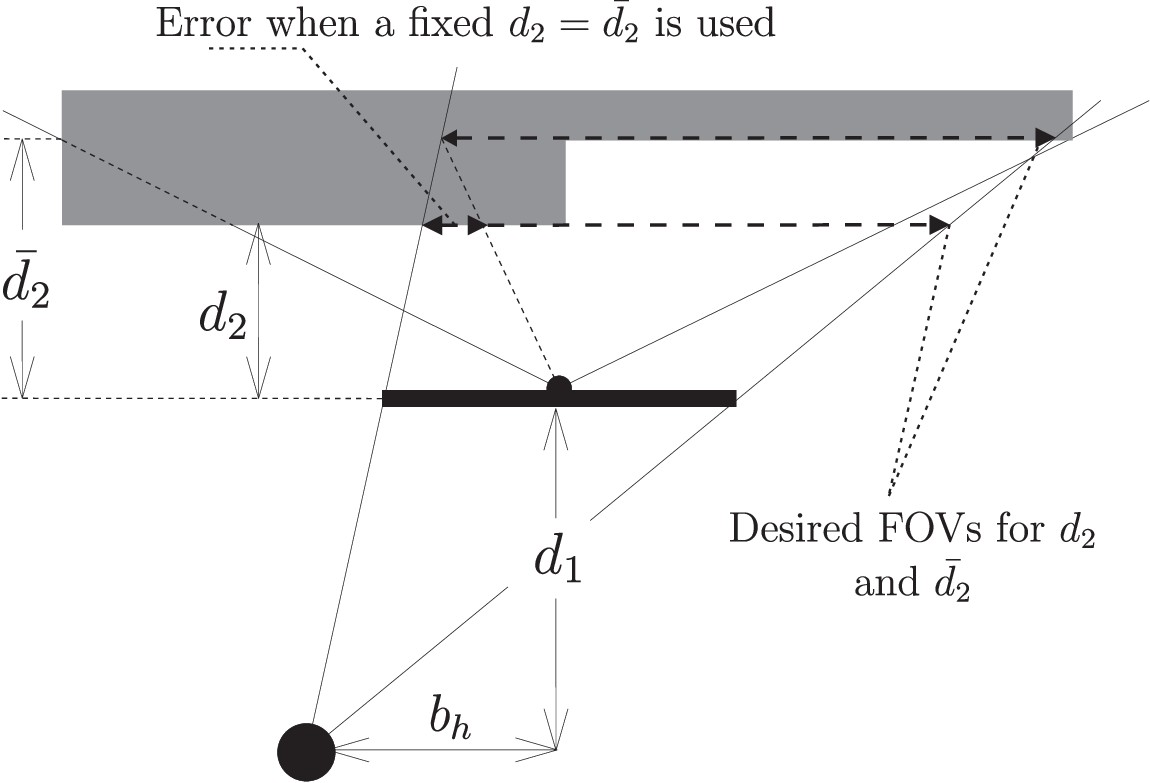}
\caption{Errors due to different values of $d_2$.}
\label{fig:error_d2_difference}
\end{figure}

Besides errors caused by the imprecise estimation of the parameters, another important source of errors comes from the fact that in practice the scene geometry is hardly similar to a plane. In this case, the first thing that becomes apparent is the fact that a single working distance $d_2$ is not even defined since it is different for each object on the scene, as depicted in Figure~\ref{fig:error_d2_difference}. Some methods try to address this issue by estimating a depth map for the scene and applying the compensation for each pixel of the image based on the respective distance estimated for that point \cite{Unuma14,Baricevic12}.

Besides leading to a substantial increase on the computational cost and inserting gross distortions on the image in the case of errors in the depth map estimation, these methods still fall short of addressing the effects caused, for example, by occlusions caused by depth discontinuities.
We propose to model the distortions originated from the scene geometry in the planar geometry methodology through the study of three simpler effects: the error due to the difference in the working distance for each point in the image, the error for occluded surfaces that occur when certain areas of a scene are hidden behind an object, and the error for foreshortened surfaces, which consists of surfaces that appears to be smaller when viewed through an oblique angle.
The latter two effects constitute a problem for a see-through system because, being view-dependent, occluded or foreshortened surfaces will be differently perceived trough the user's and by the \textit{Frontal Camera}'s viewpoint.

Evaluating the errors caused by different working distances for each point in the scene is somewhat straightforward. Considering a fixed working distance chosen as $\bar{d_2}$, we can perceive the distance fluctuations around $\bar{d_2}$ for the remaining points of the scene as errors in the estimation of $d_2$. This way, the error between the desired and the camera view for each position on the scene will be given by the error in the working distance estimation on that point, weighted by the model sensitivity $\phi_{d_2}(\bar{d_2})$ which was evaluated earlier in \eqref{EQ:ANALYSIS_D2}. The conclusions reached previously can be directly generalized for this situation, with the exception that the error analysis is localized and thus different for each position, instead of global for the whole scene. The most important conclusions that are reached from the previous analysis is that, for large enough working distances $d_2$, errors caused by different object heights can be neglected and that larger angles of view $\theta_F$ reduce the system sensitivity and consequently improve the image quality for the approximations adopted.

\begin{figure}[ht]
\centering
\includegraphics[width=80mm]{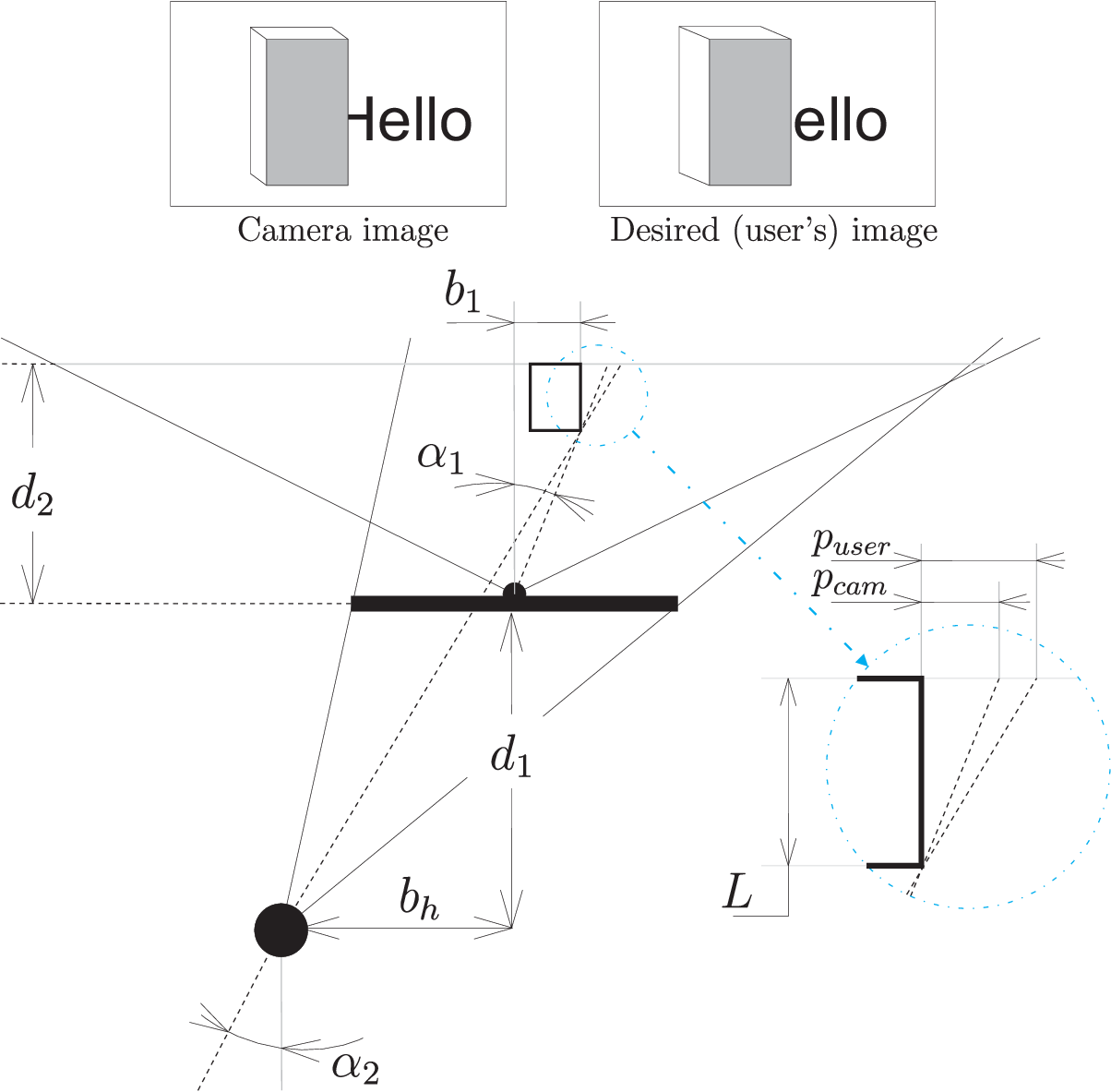}
\caption{View dependence effect of occlusion.}
\label{fig:occlusion}
\end{figure}

In studying the effects of occlusion in a see-through system, the view dependence effect on the length of an occluded surface is modelled trough a scene with a regular hexahedron occluding object as depicted in the diagram of Figure \ref{fig:occlusion}. Since the position of the user and camera are different from each other, the width $p_{\{user,cam\}}$ of the area occluded by the object is different for each point of view.

In order to express how much the occluded surface in the acquired image matches that of the desired view, we compute the ratio $p_{cam}/p_{user}$ between the occluded surfaces corresponding to the viewpoints of the user and \textit{Frontal Camera} (representing the desired and the acquired images, respectively). This can be done using trigonometric relations as:
\begin{eqnarray} \nonumber
    { \frac{b_1+p_{cam}}{d_2}=\frac{p_{cam}}{L}}
    &\!\! \text{,} &
    \!\!\ { \frac{b_h+b_1+p_{user}}{d_1+d_2}=\frac{p_{user}}{L}} \\
    \nonumber
    p_{cam}=\frac{Lb_1}{d_2-L} 
    &\!\! \text{,} &
    \!\!\ p_{user}=\frac{L(b_h+b_1)}{d_1+d_2-L}
\end{eqnarray}
\begin{equation} \label{EQ:OCLUSION}
    \frac{p_{cam}}{p_{user}} = \frac{b_1(d_1+d_2-L)}{(d_2-L)(b_h+b_1)}
    \,\text{.}
\end{equation}

In order to achieve a more intuitive expression for equation \eqref{EQ:OCLUSION}, we are first going to make some simplifications based on the expected values for the working conditions. If we assume that the occluding object is not too close to the obstacle relative to the scene ($d_2\gg L$), and that the scene distance is much larger than the user-obstacle distance ($d_2\gg d_1$), we can approximate $p_{cam}/p_{user}$ by

\begin{equation} \label{EQ:OCLUSION2}
    \frac{p_{cam}}{p_{user}}\approx\frac{b_1 d_2}{(b_h+b_1)d_2}=\frac{b_1}{b_h+b_1}
    \, \text{.}
\end{equation}

It can be seen that under the aforementioned hypothesis the ratio $p_{cam}/p_{user}$ is close to $1$ as long as $b_1\gg b_h$, which corresponds to either the user being close to the focal axis ($b_h$ small) or to the occluded surface being far from the center of the obstacle, giving a satisfactory match between the occlusion perceived by the desired and the \textit{Frontal Camera} views.
On the other hand, a relatively small $d_2$ (where $L<d_2$) can lead to substantial errors. In this case, if we assume that $b_h$ is very small, $p_{cam}/p_{user}$ can be expressed as
\begin{equation} \label{EQ:OCLUSION3}
    \frac{p_{cam}}{p_{user}}\approx\frac{b_1(d_2-L)+b_1d_1}{b_1(d_2-L)}=1-\frac{d_1}{d_2-L}
    \, \text{.}
\end{equation}

It can be seen that the second term of the last equation, for small $d_2-L$ compared to $d_1$, leads to a significant error in the ratio of the occluded surfaces, which implies that under this setting the see-through system will be significantly influenced by occlusions.
Nevertheless, when the scene distance is large the absolute values of the occluded dimensions $p_{cam}$ and $p_{user}$ generally get smaller, which means that the absolute difference of the occlusions between both views will get less noticeable (although their ratio is different from $1$).
This behavior will be later illustrated through computer simulations. It is also interesting to point out that the occlusion error is independent of the parameters $D$ and $\theta_F$.

\begin{figure}[ht]
\centering
\includegraphics[width=8cm]{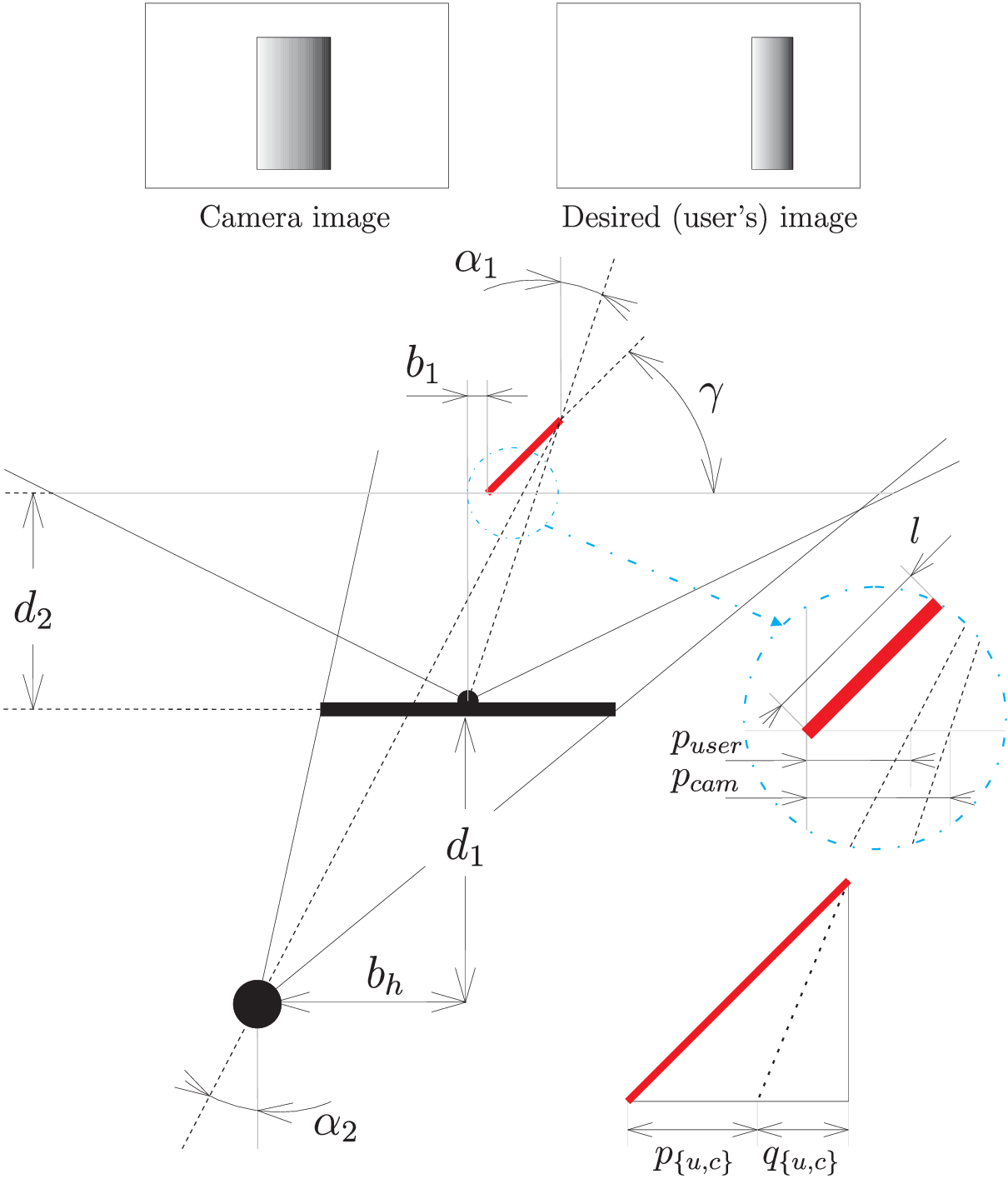}
\caption{Diagram illustrating the foreshortening effect, where $p_{\{u,c\}}$ and $q_{\{u,c\}}$ stands for $p_{\{user,cam\}}$ and $q_{\{user,cam\}}$.}
\label{fig:foreshortening}
\end{figure}

The view dependence effect of the foreshortened surfaces will be evaluated using a simple model consisting of an angled object as depicted in Figure~\ref{fig:foreshortening}. The distortion of the dimension will be evaluated through the ratio between the length of the angled surface as observed through the camera and user's viewpoints.
Since the object is not parallel to the display, its perceived width will be the projection $p_{\{user,cam\}}$ of the surface length $L$ into an imaginary plane set on the object's corner, which was assumed to be positioned at the scene distance $d_2$.

The projected object's width corresponding to the views of the user and \textit{Frontal Camera} is given by $p_{user}$ and $p_{cam}$, respectively. By using the following trigonometric relations:
\begin{equation}
\alpha_1=\arctan\left(\frac{b_1+l\cos(\gamma)}{d_2+l\sin(\gamma)}\right)
\,\text{,}
\end{equation}
\begin{equation}
\alpha_2=\arctan\left(\frac{b_1+b_h+l\cos(\gamma)}{d_1+d_2+l\sin(\gamma)}\right)
\,\text{,}
\end{equation}
\begin{equation}\label{FORESH2}
p_{\{user,cam\}}+q_{\{user,cam\}}=l\cos(\gamma)
\,\text{,}
\end{equation}
\begin{equation}
\tan\left(\frac{\pi}{2}-\alpha_{\{1,2\}}\right)=\frac{l\sin(\gamma)}{q_{\{1,2\}}} \, \text{,}
\end{equation}
it is possible to show that the ratio between the desired and the displayed foreshortened width $p_{cam}/p_{user}$ will be given by:
\begin{align} 
    \nonumber
	\frac{p_{cam}}{p_{user}} {}={}& \frac{d_1+d_2+l \sin(\gamma)}{d_2+l \sin(\gamma)} \\
    \label{EQ:FORESH1} 
    &\cdot \frac{d_2 \cos(\gamma)-b_1 \sin(\gamma)} {(d_1+d_2) \cos(\gamma)-(b_h+b_1) \sin(\gamma)}
	\,\text{.}
\end{align}

Although the complexity of the expression in \eqref{EQ:FORESH1} precludes an intuitive and complete analysis, it is still possible to attain important conclusions regarding its behavior by exploiting the symmetry present in the equation.
Excluding some trivial cases like $\gamma=0$, it can be seen that the ratio $p_{cam}/p_{user}$ will approach $1$ if the numerator and denominator cancel each other for each of the two individual fractions in equation \eqref{EQ:FORESH1}. In order for this to be satisfied, it is necessary that $d_1+d_2\approx d_2$ and $b_h+b_1\approx b_1$, which may happen either when both $d_2\gg d_1$ and $b_1\gg b_h$ for a moderate scene distance, or when $d_2$ is large enough such that $d_2\gg\{d_1,b_1,b_h\}$. The latter condition can be demonstrated by evaluating the limit of the ratio:
\begin{equation}\nonumber
    \lim_{d_2\rightarrow\infty}\frac{p_{cam}}{p_{user}}=1
	\, \text{,}
\end{equation}
which shows that for large scene distances the width of the foreshortened surface observed through the viewpoint of the \textit{Frontal Camera} coincides with that observed through the user's.
On the other hand, if either $b_h\sim b_1$ or $d_1\sim d_2$, which may happen for small values of $d_2$, significant discrepancies between the surface widths for the two observers can be noticed. This behavior will later be illustrated in a computer simulation. This shows that the view dependence characteristic of foreshortened surfaces, although possibly significant, becomes negligible with the increase of the scene distance.

It is important to notice that on the analysis made regarding both occlusion and foreshortening, the view dependent distortions were independent of the angle of view of the \textit{Frontal Camera}, which enables an unrestrained choice for this parameter in order to achieve the best performance of the system.
Furthermore, it was shown in the present evaluation that the scene distance have a great impact on the performance of video see-through systems. The planar geometry based systems studied are extremely benefited by working conditions employing large values of $d_2$, to the point of allowing the use of drastic approximations that may result in important computational savings with little impact on the system's performance.

\section{Results} \label{SIMUL}
This section is divided in two parts. First, the results obtained in Section \ref{ANALYSIS} are verified numerically through computer simulations in order to study the expected system performance in some scenarios of practical interest. Afterwards, a qualitative evaluation of a planar geometry see-through system is made using an implementation in a \mbox{3-D} virtual reality simulator with a driver assistance system as a sample application, illustrating its performance under the proposed $d_2\rightarrow\infty$ approximation for the scene distance as well as the influence of occlusion and foreshortening. A physical implementation finally illustrates the system working on a real environment. Throughout this section, all distances are assumed to be given in meters.

\subsection{Simulation of Parameter Behavior}
\begin{figure}[ht]
\centering
\includegraphics[width=9cm]{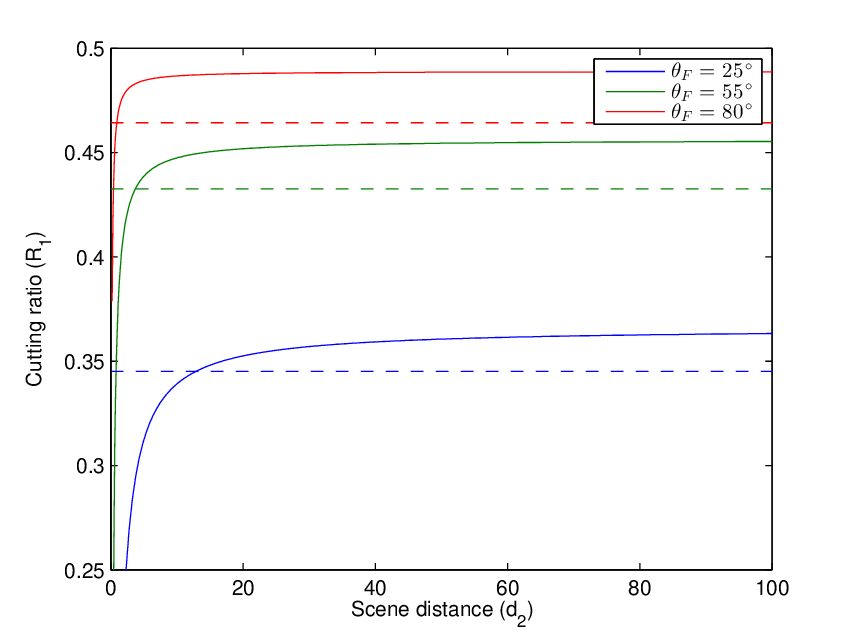}
\caption{Left cutting ratio as a function of $d_2$ and $\theta_F$.}
\label{fig:simRatio}
\end{figure}

In order to quantitatively assess the behavior of the distortions in the proposed see-through system, the results predicted in the previous sections will be evaluated numerically for some working conditions.
First, the evolution of the cutting ratio given by equation \eqref{EQ:RATIO1} in function of the scene distance $d_2$ and angle of view $\theta_F$ will be considered. The errors due to the proposed approximation $R(d_2):=R(\infty)\,\forall d_2$ will be accounted for by considering the difference between the real cutting ratio for a given working distance $R_{\{1,2\}}(d_2)$ and the fixed ratio $R_{\{1,2\}}(\infty)$.
We also illustrate the influence of a lower bound on the accepted performance by checking the intervals where the real $R(d_2)$ is reasonably close to the fixed $R(\infty)$, thereby depicting a set of admissible working distances. The resulting cutting ratios for parameters $d_1=1$, $b_h=0.125$ and $D=0.5$ is depicted in Figure \ref{fig:simRatio}, with the bound set as $0.95R(\infty)<R(d_2)<1.05R(\infty)$ in the dashed line (the upper value is omitted in the plot since $R(d_2)$ increases monotonically).

It can be seen that the larger the value of $\theta_F$, the faster $R(d_2)$ converges to $R(\infty)$, which leads to a smaller scene distance being needed in order to attain the specified quality bounds for the proposed approximation. Furthermore, large $\theta_F$ leads to a smaller error $|R(d_2)-R(\infty)|$, therefore decreasing the errors/distortions due to non-planar scene geometry. For the example in Figure \ref{fig:simRatio}, with a value of $\theta_F=80^{\circ}$ we have that $R(d_2)>0.95R(\infty)$ is met for all $d_2>0.9$. This illustrates the importance of the choice of a large angle of view for the \textit{Frontal Camera}.

\begin{figure}[ht]
\centering
\includegraphics[width=9cm]{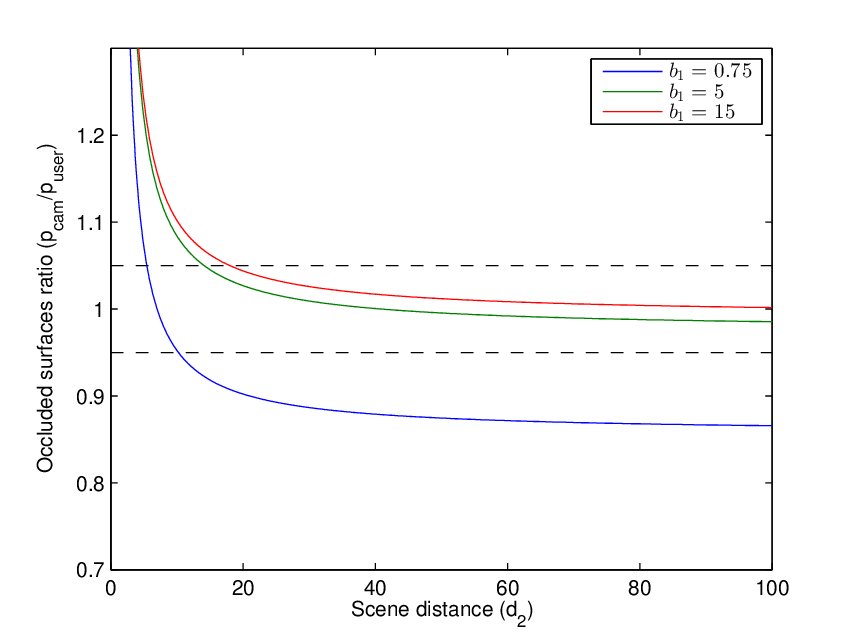}
\caption{Occluded surfaces ratio as a function of $d_2$ and $b_1$.}
\label{fig:simOcclusion}
\end{figure}

To evaluate the effect of occlusion, the ratio $p_{cam}/p_{user}$ given in equation \eqref{EQ:OCLUSION} is plotted on Figure~\ref{fig:simOcclusion} as a function of $d_2$ and $b_1$ for parameters $d_1=1$, $b_h=0.25$ and $L=1$. Like the previous simulation, we set a lower and upper bound on the computed ratio as $0.95<p_{cam}/p_{user}<1.05$ in order to illustrate an interval of acceptable performance, depicted by the dashed lines.

It can be seen that although the occluded surfaces ratio converges to a fixed value, it is generally not equal to $1$. Furthermore, as $d_2$ increases, the error $|p_{cam}/p_{user}-1|$ is larger when the distance $b_1$ is closer to the user's horizontal position $b_h$, agreeing with the results obtained previously on equation~\eqref{EQ:OCLUSION2}.
Nevertheless, if the scene distance is large compared to the remaining parameters, the absolute error $|p_{cam}-p_{user}|$ will be small, reducing the influence of errors/distortions due to occlusions on the scene.

\begin{figure}[ht]
\centering
\includegraphics[width=9cm]{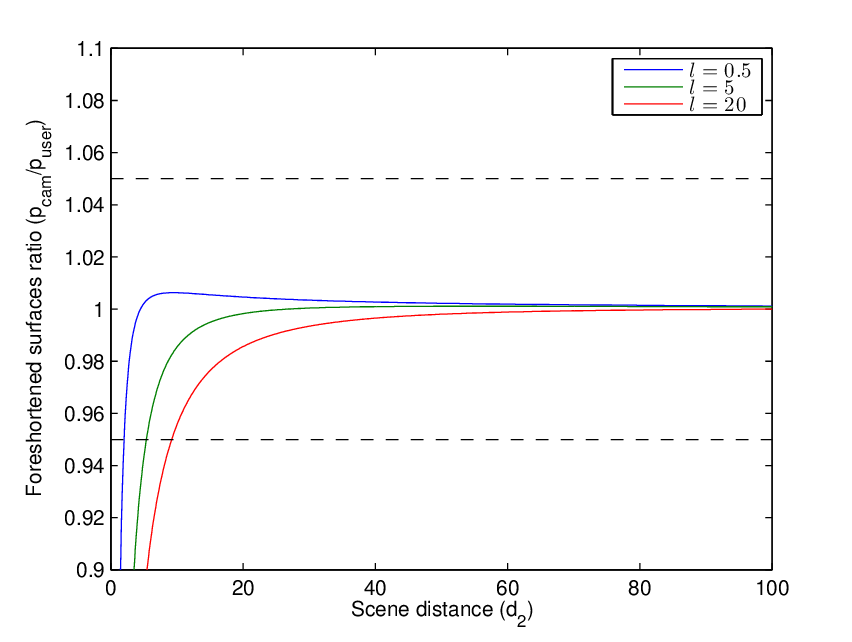}
\caption{Foreshortened surfaces ratio as a function of $d_2$ and $l$.}
\label{fig:simForeshortening}
\end{figure}

In order to illustrate the case of foreshortening, the ratio between the real and the displayed surfaces $p_{cam}/p_{user}$ in equation~\eqref{EQ:FORESH1} is plotted on Figure~\ref{fig:simForeshortening} as a function of $d_2$ and $l$, with the remaining parameters set as $d_1=1$, $b_h=0.125$, $b_1=0.25$ and $\gamma=45^{\circ}$. A lower and upper bounds are set again as $0.95<p_{cam}/p_{user}<1.05$ to illustrate intervals with acceptable performance, and are depicted by the dashed lines.

It can be seen that for large scene distances the ratio converges monotonically to $1$ with the increase of $d_2$ as predicted in the previous section.
On the other hand, for values of $d_2$ that are small compared to $d_1$, the ratio starts to deviate significantly from $1$ since $b_1$ is also of the same order of $b_h$ for this example.
It can also be noticed that large values of $l$ decrease the sensitivity of the first/left fraction of the ratio $p_{cam}/p_{user}$ in equation~\eqref{EQ:FORESH1} to variations of $d_2$, specially when the latter is small, which explains the slower convergence of the curve seen in the plot for larger $l$, although the effect it has on the ratio when combined with the remaining variables is nontrivial.

\subsection{System Implementation}

In order to provide a visual illustration of the see-through methodology studied and the proposed approximations, first a virtual reality simulation was developed in Unity 3D\textsuperscript{TM} containing the user, obstacle and a scene, providing a controlled environment to observe the solution without the need for the user's position estimation, since it was obtained by software, therefore reducing the numbers of error sources. A physical implementation of the system is illustrated later.

\begin{figure}
\centering
\includegraphics[width=8cm]{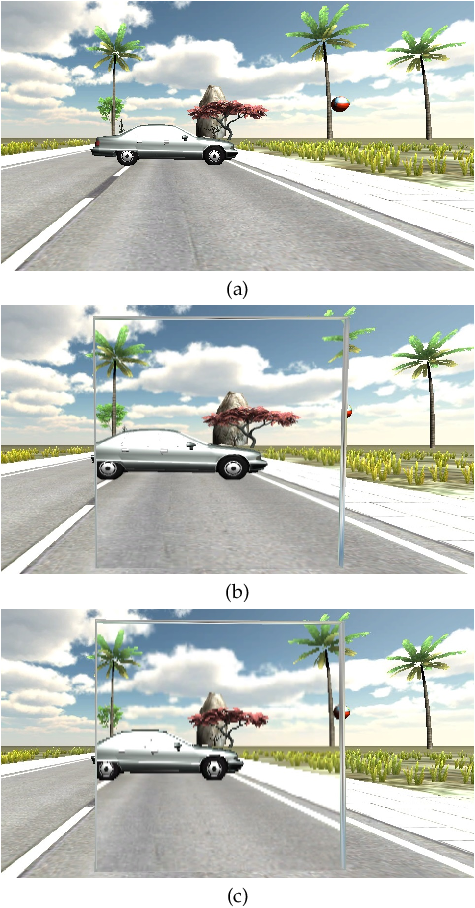}
\caption{Qualitative simulation with car close  (at approximately 4m) to the obstacle. (a) Ground truth (un-occluded) scene. (b) Results with $\theta_F=30^{\circ}$ and $D=1$. (c) Results with $\theta_F=63.5^{\circ}$. }
\label{fig:simul_car_close}
\end{figure}

\begin{figure}
\centering
\includegraphics[width=8cm]{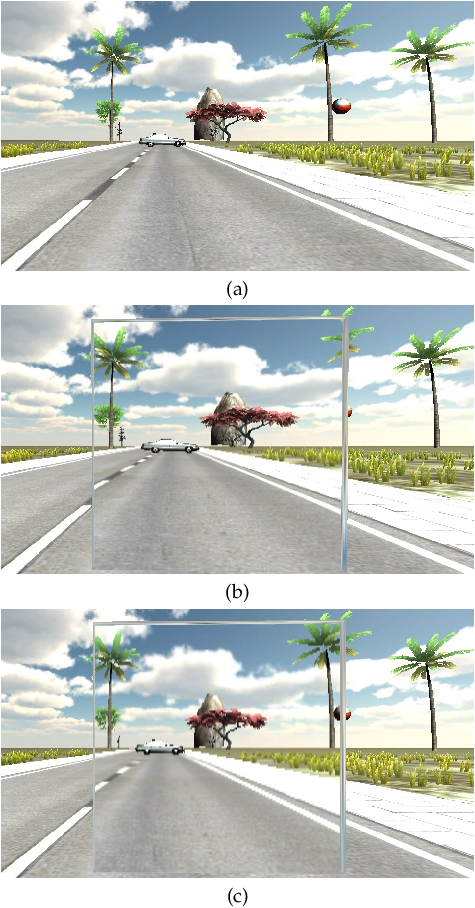}
\caption{Qualitative simulation with car far (at approximately 15m) of the obstacle. (a) Ground truth (un-occluded) scene. (b) Results with $\theta_F=30^{\circ}$ and $D=1$. (c) Results with $\theta_F=63.5^{\circ}$. }
\label{fig:simul_car_far}
\end{figure}

The simulation that was chosen as a setting is similar to that found in a driver assistance (DAS) system, which favored the evaluation of the previously studied distortions/errors due to the complexity and variability of its content.
The scene was composed of an horizontal ground plane with trees, a street, and objects like a car, a cube and a ball. The obstacle was mounted on the street at the user's view position, illustrating a situation where the driver's view would be occluded by a pillar of the vehicle or by fog. In this case conventional or near-infrared imaging can be used for image dehazing and/or disocclusion \cite{Schaul09}.
Since the display was not enclosed (i.e., it was left ``floating''), clear visualization of its edges was possible, allowing a better assessment of errors near the borders. The original scene is illustrated in Figures~\ref{fig:simul_car_close}-(a) and~\ref{fig:simul_car_far}-(a).

On the first simulation, we desire to illustrate the effect of the approximation of $R(d_2):=R(\infty)$ and the influence the angle of view have on the resulting errors. In this case, we compare the ground truth of the unobstructed scene with the see-through system image containing an object (a car) both close and far from the obstacle to illustrate the distortions experienced by different errors in the working distance due to the approximation, using two different values for $\theta_F$ (without loss of generality). The results are depicted in Figures~\ref{fig:simul_car_close} and~\ref{fig:simul_car_far}.


It can be seen that despite the complex scene with objects at many different distances, the strong approximation of $R(\infty)$ employed in order to avoid estimating $d_2$ seems to give a good result for appropriate (larger) $\theta_F$, as illustrated by the closer matching between the obstacle borders and the remaining background scene. This indicates that in some situations, neither the estimation of the working distance nor the full scene reconstruction may be necessary, leading to important computational savings.

Furthermore, the objects that are farther from the obstacle seems to experience less distortions than those which are near. This becomes apparent if we compare the errors at the borders of the obstacle for the ground and street, which are close to the camera and seems to present a larger mismatch, and the trees, ball and clouds, which are farther and display a better correspondence between their real and displayed versions.

Another important aspect is the influence of the angle of view of the \textit{Frontal Camera} in this example. For the proposed approximation, a larger $\theta_F$ apparently reduces the distortions on the reconstructed image, as it can be observed for the trees, ball and clouds. Besides the increased perceptual quality of the results in Figures~\ref{fig:simul_car_close}-(c) and~\ref{fig:simul_car_far}-(c), this also led to a better representation of the dimensions of the object, which resulted, for example, in the displayed width of the car being significantly closer to the real one when $\theta_F=63.5^{\circ}$ than when $\theta_F=30^{\circ}$, both when it was close and far from the obstacle.

\begin{figure}
\centering
\includegraphics[width=8cm]{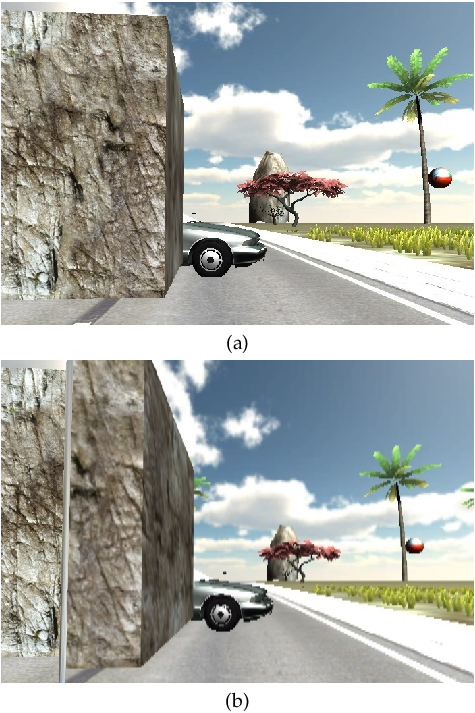}
\caption{Qualitative simulation with a foreshortened surface ($l=5$, $b_1=-1.5$, $d_2=1.3$, $d_1=2$, $b_h=0.7$, $\gamma=90^{\circ}$). (a) Ground truth (un-occluded) scene. (b) See-through result.}
\label{fig:simul_car_foreshortening}
\end{figure}

\begin{figure}
\centering
\includegraphics[width=8cm]{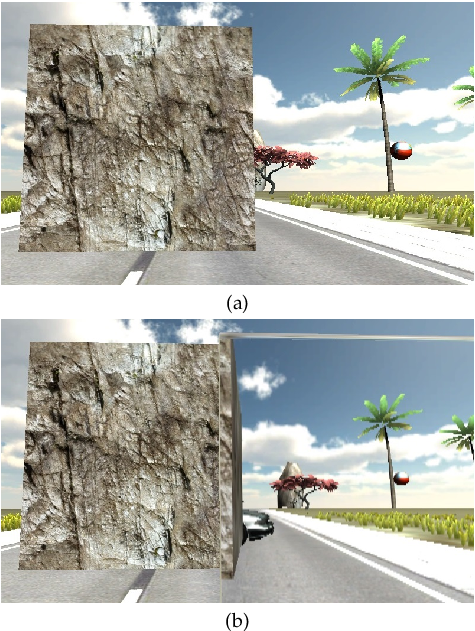}
\caption{Qualitative simulation with an occluded surface ($L=5$, $b_1=-0.35$, $d_2=6.3$, $d_1=2.5$, $b_h=1.75$). (a) Ground truth (un-occluded) scene. (b) See-through result.}
\label{fig:simul_car_occlusion}
\end{figure}

In order to visually illustrate the influence of occlusion and foreshortening in this example, two different simulations are devised with a cube-shaped object relatively close (at 1.3 meters) to the obstacle and a car partially occluded by it. The results can be seen in Figures~\ref{fig:simul_car_foreshortening} and~\ref{fig:simul_car_occlusion}.


The inspection of the figures illustrates that the effects of both occlusion and foreshortening may be noticeable when the object of interest is very close to the obstacle.
For the foreshortening case in Figure~\ref{fig:simul_car_foreshortening}, although the distortions experienced by objects farther from the display are negligible in this example, a noticeable distortion can be seen on the right surface of the cube, which has its perspective altered.
The ratio between the real and the displayed cube surface computed through equation~\eqref{EQ:FORESH1} was $p_{cam}/p_{user}=2.47$, whereas the ratio measured in the simulation was $p_{cam}/p_{user}\approx2.64$, which indicates a good match between the prediction made by the model derived in the previous section and the distortion experienced in the simulated example.

A similar behavior is observed for the case of occlusion depicted by the example in Figure~\ref{fig:simul_car_occlusion}, where a significant difference can be noticed in the displayed version of the region of the scene corresponding to the right part of the street. On the original scene the street and the car are completely occluded by the cube, whereas on the displayed image a significant part of both is visible, which generated an image that, although of good quality, does not perfectly matches the real scene.

A close correspondence between the distortion predicted through the proposed model and experienced in simulation is also observed for this example, since the ratio between the size of the real and displayed occluded surfaces computed through equation~\eqref{EQ:OCLUSION} results in $p_{cam}/p_{user}=-0.73$, while the ratio measured in the simulation was $p_{cam}/p_{user}\approx-0.71$.

The system was physically implemented on a hardware setting that consisted of a portable computer, whose webcam was used to acquire the image for the facial detection. The frontal camera consisted of an analog camera with an angle of view $\theta_F=24^{\circ}$. The facial detection was made using Viola-Jones' algorithm \cite{Viola04}. The notebook monitor was also used as the obstacle, yielding the value of $D=0.34$.

\begin{figure}[ht]
\centering
\includegraphics[width=85mm]{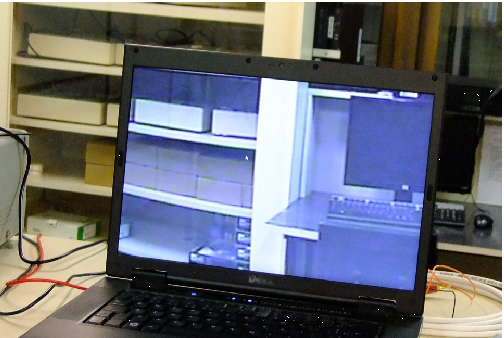}
\caption{Physical system implementation.}
\label{fig:implementation}
\end{figure}

The working system is illustrated in Figure~\ref{fig:implementation}. It can be seen that distortions occur near the borders of the projected image. They originated from both the short scene distance and the lens of the \textit{Frontal Camera}, which should present a relatively small focal distance in order to provide a larger angle of view. The algorithm, implemented in Matlab\textsuperscript{TM}, presented a runtime of $4$ seconds per frame. Considering the implementation was done in a high level programming language and that most of the time was used on the facial detection step, with an adequate platform and implementation the algorithm can easily be made to run in real time.

\section{Conclusion}
In this paper, a planar geometry based see-through methodology was studied in detail. The comprehensive description of the principles and workings of the system provided a more conclusive acquaintance regarding both the potentials and pitfalls of the technique.
A theoretical evaluation of the influence of each system parameter was performed using differential sensitivity analysis, and the effect of errors in the assumed planar scene model, like different object heights, occlusion and foreshortening, were explicitly modelled.
It was shown that for reasonably large scene distances, neither the estimation of the scene distance nor the reconstruction of its geometry are necessarily required, as opposed to what was previously believed, allowing the employment of a fixed and arbitrarily large value for the working distance. Furthermore, a large field of view for the \textit{Frontal Camera} significantly reduce errors due to the proposed approximation.
The theoretical findings were illustrated quantitatively, through the evaluation of the mathematical expressions obtained for some situations of interest, and qualitatively, through a virtual reality simulation in idealized conditions. Both results offered a good agreement with the theoretical predictions.
Finally, a physical hardware implementation illustrated the effectiveness of the proposed approach in real working conditions.

\bibliographystyle{unsrt}
\bibliography{references}

\end{document}